\documentclass[fleqn,10pt]{wlscirep}

\usepackage[version=3]{mhchem} 
\usepackage{graphicx}
\usepackage{epstopdf}
\usepackage{setspace}
\title{Nanoscale Structure, Dynamics, and Aging Behavior of Metallic Glass Thin Films} 

\author[1,2,$\dagger$]{J.A.J. Burgess}
\author[1,3]{C.M.B. Holt}
\author[1,4]{E.J. Luber}
\author[2]{D.C. Fortin}
\author[2]{G. Popowich}
\author[1,3]{B. Zahiri}
\author[1]{P. Concepcion}
\author[1,3,5]{D. Mitlin}
\author[1,2,*]{M.R. Freeman}

\affil[1]{National Institute for Nanotechnology, Edmonton, Canada}
\affil[2]{University of Alberta, Department of Physics, Edmonton, Canada}
\affil[3]{University of Alberta, Department of Chemical and Materials Engineering,  Edmonton, Canada}
\affil[4]{University of Alberta, Department of Chemistry, Edmonton, Canada}
\affil[5]{Clarkson University, Chemical \& Biomolecular Engineering, Potsdam, New York,  United States}
\affil[$\dagger$]{Now at the Max Planck Institute for Structure and Dynamics of Matter, Hamburg, Germany and the Max Planck Institute for Solid State Phsyics, Stuttgart, Germany}
\affil[*]{mark.freeman@ualberta.ca}

\begin{abstract}
Scanning tunnelling microscopy observations resolve the structure and dynamics of metallic glass Cu$_{100-x}$Hf$_{x}$ films and demonstrate scanning tunnelling microscopy control of aging at a metallic glass surface. Surface  clusters exhibit heterogeneous hopping dynamics. Low Hf concentration films feature an aged surface of larger, slower clusters. Argon ion-sputtering destroys the aged configuration, yielding a surface in constant fluctuation. Scanning tunnelling microscopy can locally restore the relaxed state, allowing for nanoscale lithographic definition of aged sections. 
\end{abstract}
\begin{document}

\flushbottom
\maketitle
\doublespacing
%
%
\thispagestyle{empty}

\section*{Introduction}

Metallic glass (MG) thin films are a novel class of materials of increasingly widespread use and interest \cite{GenMG}. Their unique combination of mechanical, electrical and topographical properties\cite{JRGreerNanolett} fosters applications in nano/micro-electrical mechanical systems\cite{Erik,luber2008tailoring,ophus2008resonance,nelson-fitzpatrick_synthesis_2007}, flexible-transparent conductors\cite{lin2013improved}, nanoimprinting\cite{Schroers2009} and biomedical coatings\cite{subramanian2013fabrication, chiang2010surface}. Additionally, they show potential as alternatives to single crystal substrates as ultra-smooth platforms for top-down patterned nanostructures. However, these surfaces are complex and incompletely understood. Deeper insight into their properties is critical for the rational design of thin film MG coatings for desired applications.

Glassy behavior is a near-universal phenomenon in highly disordered condensed matter systems\cite{GlassyRMP2011}. Dynamics in the deep-supercooled regime are of special interest as they influence the properties of disordered materials across all relevant time scales\cite{zhao2013,ferrante2013acoustic}. At temperatures below the glass transition, MGs are believed to be composed of clusters of constituent atoms\cite{Chen2015}, which are densely packed, often possessing some degree of medium range ordering\cite{Sheng}. Well below the glass forming temperature, aging is dominated by collective excitations\cite{AdamandGibbs} of the cluster units\cite{BetaSLCAging}. Many of the properties of MGs are described theoretically in the context of models built on the dynamics of these clusters\cite{chen_mechanical_2008}. Of particular interest is the manifestation of heterogeneous dynamics\cite{RichertPRL2011} and the effects of geometrical constraints\cite{NatMatwall}. Constraints such as surfaces can accelerate or slow dynamics\cite{RichertReview}. Recent results on film surfaces indicate accelerated aging to an equilibrium state having slow dynamics\cite{AccSurf}. Scanning probe investigations have shown slow cluster hopping\cite{PCL}, interpreted as isolated secondary ($\beta$) relaxations\cite{SamwerMatToday}, in  MGs\cite{PCL} and amorphous silicon\cite{PRLSiHop}. A complementary approach, dynamic force microscopy, has been used to measure nanoscale heterogeneity by measuring dissipation in dynamics of metallic glass surfaces\cite{DynamicFMHetero}. Resolving evolution of surfaces in real space is critical to understanding the dynamics that dictate the material structure, as has been highlighted by recent atomic force microscopy  imaging of growth mechanisms on crystalline surfaces\cite{CryGrowSci2014}. 

\section*{Results and Discussion}

In this work, the aging process of a metallic glass thin film is spatially resolved and controlled with a scanning tunnelling microscope (STM). The binary phase diagram of copper and hafnium features low substitutional solid-solubility and a large number of intermetallic compounds\cite{CuHfphase}. It is one of the few binary metallic alloys capable of glass formation in the bulk\cite{Duan}. The glass forming region is expected to be between $\sim 30-70$ at$.$\% hafnium for samples prepared via melting spinning (cooling rate $\sim 10^3 – 10^7$ K/s)\cite{CuHfGlassTrans}. Here STM observations on Cu$_{100-x}$Hf$_{x}$ are presented. The surface is found to have hillock features with sizes dependent on Hf concentration, and pervasive dynamics manifested as hopping clusters. Lower hafnium concentration films exhibit an aged surface state with slower dynamics and larger clusters. This surface state can be disrupted by either moving material on the surface with the STM tip, thus exposing the subsurface clusters, or via \em in situ \em argon sputtering. The aged surface can be locally restored via STM scanning, enabling lithographic definition of the surface morphology through accelerated aging of an unstable amorphous surface state. 

The samples were imaged in a Createc ultra high vacuum STM operated at a base pressure of $2\times 10^{-10}$ mbar. Electrochemically etched tungsten tips were used after \em in situ \em electron bombardment to remove oxide, and subsequent field emission to reduce the tip radius. Samples in five concentrations of Cu$_{100-x}$Hf$_{x}$ ( $x=15$, $x=19$, $x=25$, $x=37$, and $x=50$) were prepared by co-sputtering a 50 nm layer of CuHf onto a Si(100) wafer in a high vacuum deposition system with a base pressure ranging from $1\times10^{-8}$ to $8\times10^{-8}$ mbar. Samples were transferred to the STM under a clean, inert atmosphere. 

\begin{figure}[h!]
\begin{center}
\includegraphics[scale = 0.8]{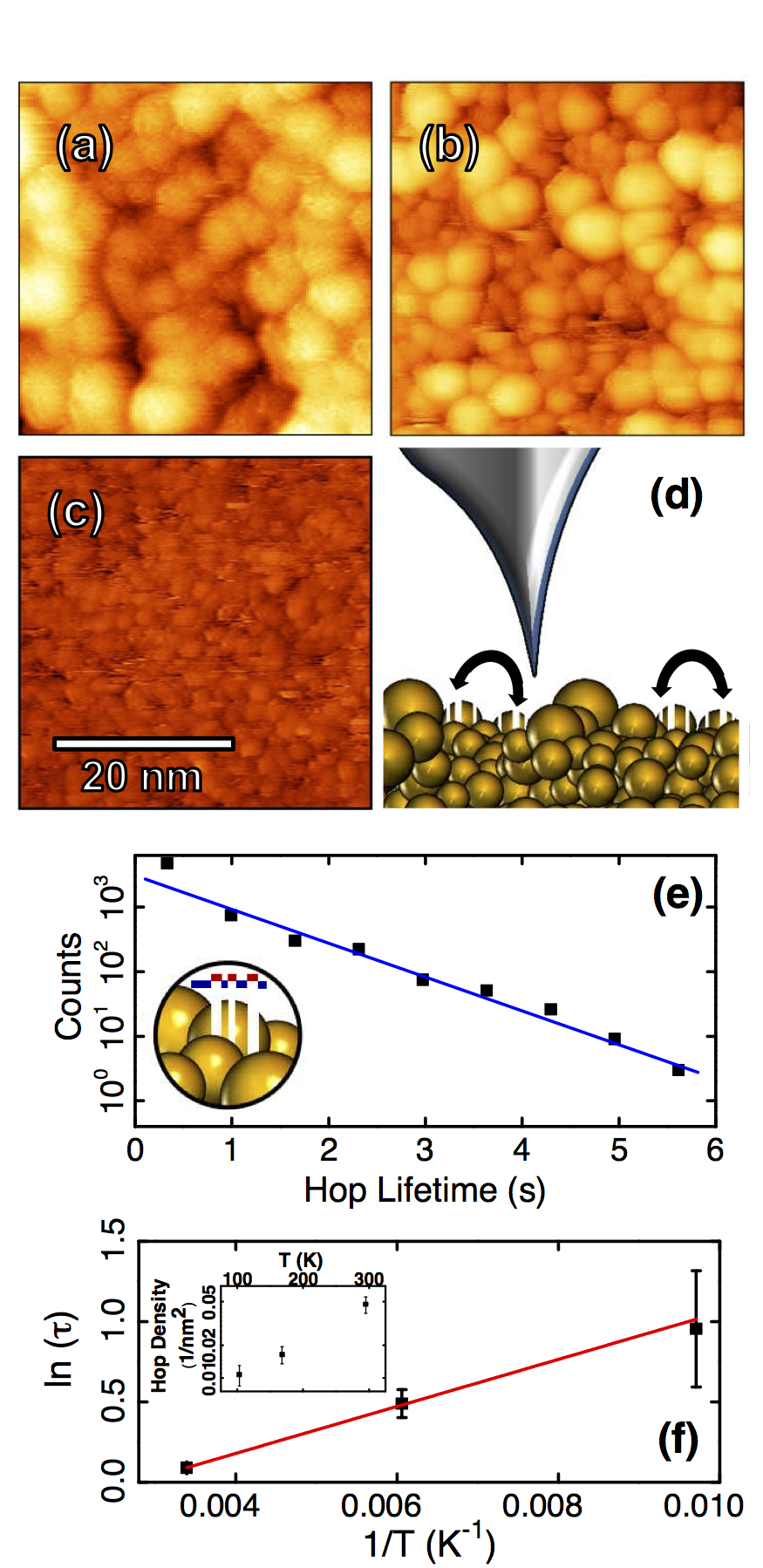}
\end{center}
\caption{\label{fig:Struct} Topographic images of films with variable Hf concentration: (a) Cu$_{85}$Hf$_{15}$ (19 pA, 1.7V), (b) Cu$_{81}$Hf$_{19}$ (20 pA, 2.6V) (c) Cu$_{50}$Hf$_{50}$ (20 pA, 2.2V). All images are shown with a linear colour scale corresponding to 3$\thinspace$nm height difference between dark brown and light yellow. Small cluster substructure is visible in (b) and (c). On close inspection, some hopping clusters are partially imaged over a few scan lines (b) and (c), leading to a striped appearance, as depicted schematically in (d). Hopping can be more clearly identified in movies (see Supplementary Movies 1 and 2 online). Extracting the widths of the telegraph noise-like stripes, hopping lifetimes of the clusters may be measured (inset (e)). Using data acquired on a Cu$_{85}$Hf$_{15}$ sample (150 pA, 1.5V) the lifetime distribution is found to be exponential (e) and temperature dependent (f). The areal density of hopping sites increases exponentially with temperature (inset (f)) and suggest energy barriers of several 100 meV.} 
\end{figure}

All samples exhibited surfaces consisting of round hillocks (Fig. \ref{fig:Struct}), similar to previously studied amorphous materials\cite{MayrTipSim, Mayr, Burgler1999}. Samples with low Hf concentration ($x=15$, $19$, $25$) showed two distinct populations of hillocks: one larger sized set defining the highest points of the surface, with a smaller sized second set visible below, in gaps between upper layer hillocks (Fig. \ref{fig:Struct}b). By contrast samples with Hf concentration clearly above the nominal glass forming level ($x=37$, $50$) featured smaller less clearly delineated hillocks in a single smooth layer (Fig. \ref{fig:Struct}c). Size distributions were calculated (see Supplementary Figure 1) from images of each sample using a watershed based algorithm\cite{Gwyddion, Klapetek2003} with additional height thresholding to remove partially occluded hillocks and a correction for the non-zero tip radius\cite{TipRadCorr}. The feature size of the Cu$_{63}$Hf$_{37}$ and Cu$_{50}$Hf$_{50}$ samples as well as the second layer features in the lower Hf concentration samples agree with expected cluster size of 1-2 nm in MGs\cite{SW2010, chen_mechanical_2008, TEMabCry}. Additionally these smaller structures exhibit a packing pattern consistent with that expected for a glass\cite{Sheng}. Previous work has shown that MG thin films of Cu$_{100-x}$Hf$_{x}$ with as little as 10 at. \% Hf can be produced via co-sputtering\cite{Erik}. due to the extreme effective quenching rates of sputtering ($> 10^{15}$ K/s)\cite{SputterQuench}. This is consistent with the STM observations presented here, where all films  have an underlying amorphous cluster structure, but the surface sensitivity of the STM allows identification of  a surface reconstruction with larger clusters on samples with Hf concentration below $x=25$ feature.

STM investigation revealed stochastic dynamics of individual hillock structures (see Supplementary Movies 1 and 2 online). These dynamics indicate the hillocks are clusters of various sizes, and the dynamics are highly reminiscent of the hopping clusters observed on other MG systems\cite{PCL} and on amorphous silicon\cite{PRLSiHop}. Contrasting with previous observations, activity on Cu$_{100-x}$Hf$_{x}$ surfaces was pervasive, affecting the a significant proportion of the clusters over all time scales accessible by scanning (ms to minutes). On the lower concentration samples, hopping was primarily limited to particular sites, often in gaps between larger hillocks. Many hopping sites showed multiple shifts, indicating a flickering mechanism with semi-reproducible hopping (Fig. \ref{fig:Struct}d).  Typically, smaller sized features ranging between 1-5 nm in diameter dominated the dynamics. Larger features were observed to appear or disappear spontaneously, but not to flicker repeatedly. This clearly demonstrates the nanoscale heterogeneous dynamics that is expected on an amorphous surface. On the $x=37$ and $50$ samples, hopping was much more prominent, resulting in wide spread flickering noise. Movies showing the flickering and hopping processes are available in the Supplementary Information online. The surfaces are too active to allow tracking of individual clusters. We therefore cannot determine if individual clusters hop between two sites in reproducible two state dynamics. However, each hopping site on the surface acts as a two level system, exhibiting occupied or unoccupied states.

Extracting information about hopping clusters inherently requires analysis of STM scan noise, which requires careful consideration of spurious noise sources. In this work hopping noise is identified rigorously by demanding that, for a hop to be counted, a portion of the hopping cluster must be imaged. Therefore each event included as contributing to the glassy dynamics involves correlations within the image across multiple scan lines, and consistency between forward and backward scans in order to reject feedback effects near steep topography. These discrimination conditions reject the sources of transient noise unrelated to the surface physics of interest, as well as for surface dynamics too fast to be conclusively attributed to clusters. 

It is equally important to ensure that identified hopping events are not a result of extraneous surface adsorbates which could mimic the motion of clusters. The dynamics observed here persisted on all samples with a strong dependence on Hf concentration. Partial images of hopping clusters appear identical to static hillock structures. On films with identical Hf concentration, but capped with a 10$\thinspace$nm Au layer, we observed broad hillocks and no hopping clusters. Dynamics persisted through heating and sputtering surface treatments confirming that extraneous adsorbates are not responsible for the observed surface structure or dynamics. In complementary TEM measurements, also using the same Hf concentrations, stochastic motions of clusters were observed. Additional details on identifying hopping clusters, testing for contaminants and TEM investigations are available in the Supplementary Information online.

To investigate the nature of this hopping, we applied a second STM featuring variable temperature control (RHK UHV-3000). Scanning with a fixed tunnelling current (150$\thinspace$pA) and measuring the dwell time of clusters on the surface (Fig. \ref{fig:Struct}e), hopping activity observed on a Cu$_{85}$Hf$_{15}$ sample was found to strongly dependent on temperature (Fig. \ref{fig:Struct}f). It is important to note that the usage of a limited bandwidth technique such as STM to observe an ensemble of hopping clusters selects a sub-population of clusters measured at any given temperature. The lifetime distribution therefore predominantly represents the lifetimes of a constrained population with approximately equal hopping energy barriers. In conjunction with the narrow range of lifetimes accessible, this leads to an apparently exponential lifetime distribution, as would be expected for homogeneous dynamics. Here heterogeneity is only detectable in the lifetimes by changing the temperature to select different cluster sub-populations. 

The effects of limited bandwidth on measuring cluster dynamics have been noted in past STM studies of hopping surface clusters\cite{PCL} including the fact that they  preclude a standard Arrhenius analysis. As an alternative, we apply here  a bandwidth corrected\cite{TLSBW} analysis to each data point assuming an attempt frequency of 1 THz and find observed barriers vary between 0.25 eV at 100$\thinspace$K up to 0.67$\thinspace$eV at room temperature. This is within the range expected for collective excitations in a metallic glass\cite{BetaSLCAging}. Perhaps more meaningfully, the number of observed hopping events also depends exponentially on the temperature. The hopping barrier extracted at room temperature represents contributions from the largest sub-population of clusters, and is therefore the best estimate of an average energy barrier for hopping. Further details on temperature dependent measurements are included in the Methods section.

\begin{figure*}[h!]
\begin{center}
\includegraphics[scale=0.8]{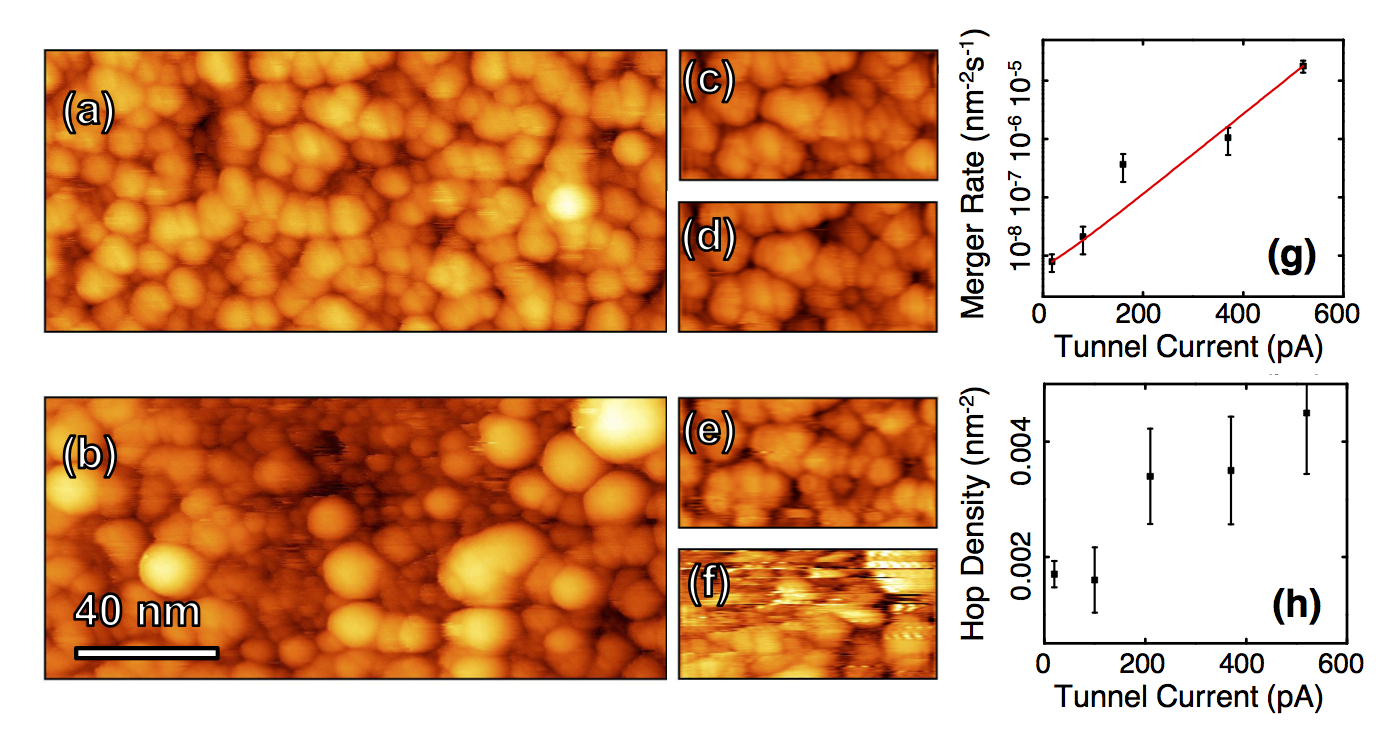}
\end{center}
\caption{\label{fig:TCladder} The effect of tunnel current on the 75/25 sample. (a) The initial scan (19 pA,2.6 V) and (b) final scan (19 pA,2.6 V) show extensive surface modification with the formation of numerous large clusters and the movement of material away from the central scan area, exposing the subsurface small clusters. Images (c)-(f) show snapshots from the tunnel current ladder with currents 100 pA, 210 pA, 370 pA and 520 pA respectively (all at 2.6 V). Each sequence consisted of 5 images acquired over 3.5 minutes. Figure (g) shows the frequency of spontaneous surface cluster mergers as a function of tunnel current as calculated over available data from the Cu$_{75}$Hf$_{25}$ sample. Figure (h) shows the counted density of highly active (flickering) clusters present on the surface during tunnel current variation experiments.}
\end{figure*}

In addition to the pervasive hopping, clusters merged irreversibly into larger features (Fig. \ref{fig:TCladder}a-f), often with no corresponding tip change. We interpret these changes as resulting from transient tunnel current spikes associated with hopping. When a cluster hops under the tip causing a large current spike, the current induced heating may be sufficient to cause local crystallization. Local crystallization in metallic glasses has been seen in TEM measurements\cite{TEMCry} and nano-indentation\cite{SciNanoIndentCry}. On other surfaces, cases of cluster formation\cite{GeCluster} and alloying\cite{STMAlloy} operating by this type of mechanism have been seen in previous STM experiments. This thermally activated change would depend exponentially on the energy transferred in the tunnel current spike, which in turn, depends linearly on the tunnel current set point, leading to a net expected exponential dependence of the merger rate on tunnel current. This is observed in the experiment (Fig. \ref{fig:TCladder}g). 

Panels \ref{fig:TCladder}c-f show images from a tunnel current ladder sequence performed on the Cu$_{75}$Hf$_{25}$ sample during which no permanent tip changes were effected. Qualitatively, the flicker activity in the freshly exposed subsurface is also significantly higher (Fig. \ref{fig:TCladder}f). Displacement of surface material and exposure of the subsurface structure was commonly seen with large mergers on the Cu$_{75}$Hf$_{25}$ surface. These data show that the samples with lower Hf concentration (Cu$_{85}$Hf$_{15}$, Cu$_{81}$Hf$_{19}$ and Cu$_{75}$Hf$_{25}$) age more quickly into a surface with slower dynamics and larger clusters.

In contrast with the mergers, the areal density of hopping sites features a weak tunnel current dependence that can be clearly identified for currents of 200 pA and above (Fig. \ref{fig:TCladder}h). To mitigate tip interaction most data was acquired at tunnel currents well below 200$\thinspace$pA and at relatively high bias voltages. This contrasts with the thermal dependence of the hopping, indicating that while the tip has an influence on hopping events, the influence is weak and secondary in comparison to the intrinsic thermal activation. The extremely active nature of MG surfaces shown here underscores the importance of understanding how these surfaces change over time if they are to be utilized as surface coatings in technological applications.

\begin{figure}[h!]
\begin{center}
\includegraphics[scale=0.8]{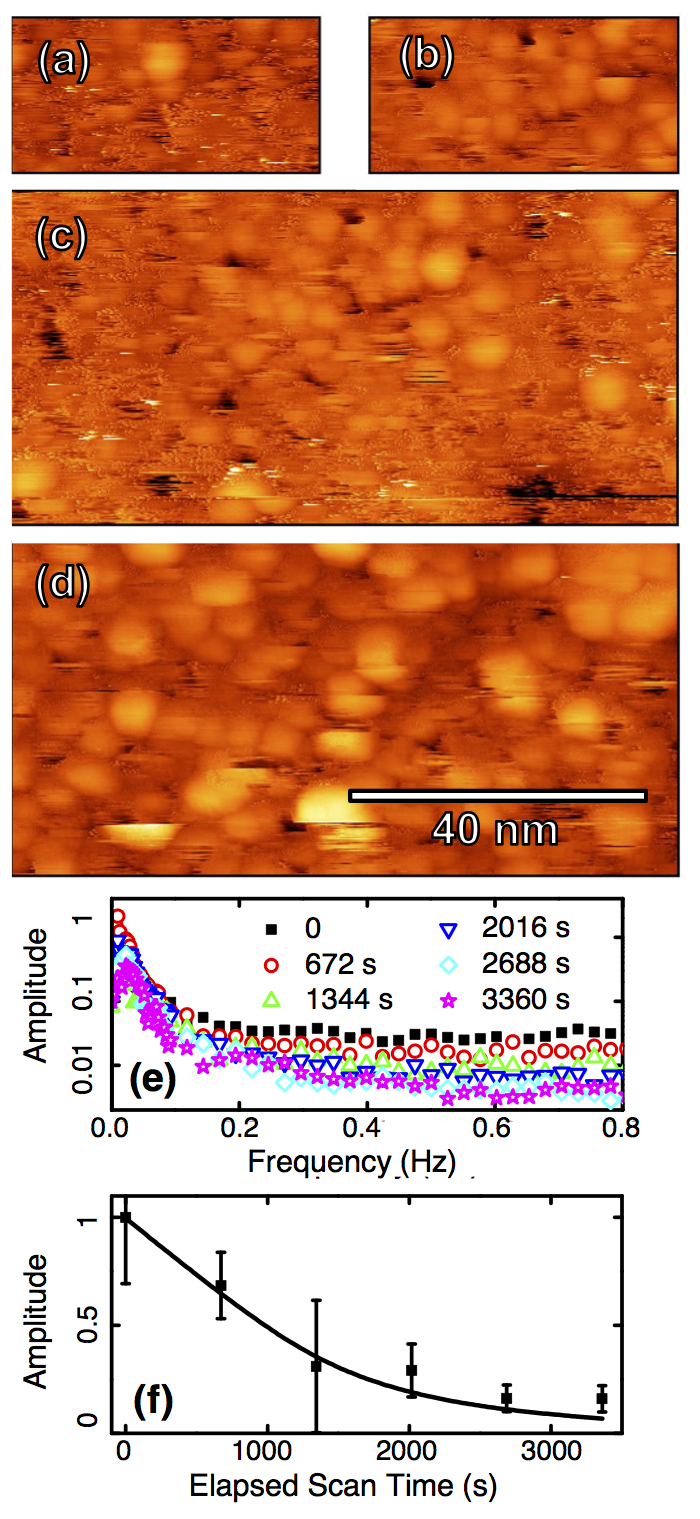}
\end{center}
\caption{\label{fig:Agingt}  At top, (a) the first scan on the re-sputtered 75/25 surface is shown along with a scan captured of the same area 22 minutes later, (b). The previously scanned region is clearly visible nested within a zoomed out unscanned region (c). At bottom (d), the same area is shown after an additional 10 scans over 2.5 hours. The surface has now stabilized in clusters. A 2.5$\thinspace$nm linear colour scale was used in all images (dark brown to light yellow). All images were acquired at 20$\thinspace$pA and 2.2$\thinspace$V. In (e), six power spectra taken in the slow scan direction show the change in the noise spectrum as the surface ages. At right (f) the noise floor extracted from each curve in (e) is fit with the calculated thermally activated noise at f = 0.5 Hz using a decaying fictive temperature ($E_B$=0.8$\thinspace$eV, $A_o$ = 1$\thinspace$THz, $\tau$=3000$\thinspace$s, $T_f(0)$=355$\thinspace$K).} 
\end{figure}

Photoexcitation has recently been shown to quicken the surface dynamics of amorphous silicon carbide\cite{Nguyen2015}.  Changes in the dynamics and structure of the surface induced by the STM tip suggest the possibility of locally controlling the surface state. We investigated this using Ar$^{+}$ sputtering of the Cu$_{75}$Hf$_{25}$ sample to disrupt the relaxed surface state. Following sputtering, the sample was immediately transferred back to the STM for inspection.  The same tip was used to take reference images just before and just after. The first measurmeents were performed approximately 10 minutes after sputtering. The surface was found to be in constant flux, featuring short scars in the fast scan directions. Over multiple scans the surface coalesced into the familiar more stable topography observed prior to sputtering. Single pixel scars were replaced by multiple scan line width truncated images of clusters and cluster flicker, as seen on untreated surfaces. Increasing the scan range, however, revealed that it was only the scanned area with stable topography and slow dynamics (Fig. \ref{fig:Agingt}). This process was repeatable over multiple sequences acquired over two days. The lithographically defined sections were found to be stable on a long ($>$24 hour) time scale with no noticeable stabilization occurring outside the scanned areas. The emergence of identifiable hopping dynamics along with the enlarged surface clusters during scan sequences confirms that the surface was driven back to an unrelaxed state in the sputtering process and the subsequent effect can be interpreted as an accelerated aging process during which the surface was restored, and hopping dynamics slowed.

A more quantitative presentation of the change in dynamics can be made by extracting the power spectrum in the slow scan direction over successive scans (Fig. \ref{fig:Agingt}e). The low frequency peak in the spectra contains topographic information, while, at higher frequency, the noise floor is dominated by jumps. The evolution of the dynamics is shown in figure \ref{fig:Agingt}(f) where the noise floor is plotted as a function of elapsed scan time. Assuming a thermally activated hopping model for the surface dynamics, the expected normalized telegraph noise spectrum is $S = \Lambda/(1+\pi^2f^2\Lambda^2)$ where $\Lambda=A_o^{-1}$exp$(E_B/(k_B T_f)$, $E_B$ is the energy barrier for diffusion and $A_o$ is the attempt frequency\cite{JAPTNspectrum}. A fictive temperature $T_f(t)= T+T_f(0)$exp$(-t/\tau)$ parametrizes the slowing of the dynamics as a thermalization process governed by a constant $\tau$\cite{RichertPRL2011}. Plotting $S(f,t)$ for a high frequency, the variation of the noise floor with time can be fit to data in figure \ref{fig:Agingt}(f). Selecting $E_B$ and $A_o$ from typical parameters for glass cluster excitation processes\cite{BetaSLCAging, PCL, PRLSiHop} ($E_B \sim$ 0.5 to 0.8 eV, $A_o \sim$ 0.01 to 1 THz), reasonable agreement is found for a best-fit of $T_f(0) \sim$ 340 K and $\tau \sim$ 1 hour. Precise values of these parameters cannot be identified due to the sensitivity of the fit and large number of unknowns, however the model gives a good description of the dynamics for physically reasonable parameters. A representative fit is shown in figure \ref{fig:Agingt}(f). The process of enlarging clusters with time and exposure to a disruption was verified using SEM measurements on a freshly deposited sample. Limitations in the resolution of the SEM, however preclude application of the same analysis technique developed for STM data. Additional information on SEM observations is available in the Supplementary Information online along with a movie demonstrating lithographic control of the surface change.

This analysis greatly simplifies the physical situation by assuming a homogeneous fictive temperature. Examining the data sequence, clusters become noticeably larger in scanned regions with slower dynamics. Therefore this fictive temperature parametrizes the energy barrier change with cluster size, and hence a range of fictive temperatures and evolution corresponding to the initial cluster size distribution would be expected. Additionally, the role of the tip is not entirely clear. The convolution of the tunnel current and tip-sample gap prevents conclusive rejection of heating via tunnel current spikes caused by hopping clusters. However, the absence of tip crashes, together with the low tunnel currents and high bias voltages used in imaging, provide strong support for the interpretation of the tip acting as a weak, transient geometric constraint on the film through a non-contact electrostatic interaction. A full analysis of this problem requires more STM observations of these systems and development of new analysis techniques. 

\section*{Conclusions}

In summary, Cu$_{100-x}$Hf$_{x}$ metallic glass thin films were examined with STM allowing direct observation of cluster structure in the films as well as temperature dependent cluster dynamics. Films with low hafnium concentrations (less than 25\%) exhibited a relaxed surface state with larger than expected cluster sizes on top of smaller features. Throughout the observations heterogeneous dynamics were observed in the form of preferential sites of high activity on the surface, increased activity in subsurface clusters, and most strikingly, through controlled aging of the surface. Destroying the relaxed surface state allowed lithographic definition of aged sections of the surface simply via scanning and verification of rapid surface aging to a stable surface reconstruction. Consideration of these surface dynamics will be pivotal for future design of MG thin film coatings. If these coatings are expected to provide corrosion resistance or antimicrobial action, change with the aging surface will be a key consideration. Conversely, the accelerated aging process may be potentially useful for top-down lithographic pattering of nanoscale structures.

\section*{Methods}

\subsection*{Sample Preparation}

Each wafer was prepared with an oxidizing cleaning step employing a wet etch of H$_2$SO$_4$ and H$_2$O$_2$ (3:1) for 30 minutes, resulting in a thin oxide layer, $\sim$ 10 nm, on the bare wafer surface. Extensive characterization of Cu$_{100-x}$Hf$_{x}$ films produced with this system has been conducted previously.\cite{Erik} With one exception, samples were transferred from the sputter chamber to the STM load lock under an inert gas (Ar or N$_2$) atmosphere. The Cu$_{85}$Hf$_{15}$ sample was transferred under a clean air atmosphere. In lieu of a typical load-lock bake, samples were stored under UHV conditions for $>$120 hrs to allow slow degassing while avoiding any thermally driven structural changes. The Cu$_{75}$Hf$_{25}$ sample was observed a second time after sputter sputtering the surface \em in situ\em\ using a typical Ar$^{+}$ cleaning process (Ionec IG70 sputter ion gun, $1.1$ kV, Ar pressure $5\times10^{-6}$ mbar, 20 minutes). No anneal was performed subsequent to the sputter sputtering. 

\subsection*{Visualization of glassy dynamics and aging through STM movies}

Three movies made with STM data are included in the Supplementary Information online). All movies are made using raw data with no additional processing beyond application of consistent grey scales and application of a leveling algorithm (three point plane subtraction). The frame rates of the movies are adjusted so that the relative frame rates are proportional to the relative scan times. 

Movie one shows a sequence of scans of the Cu$_{81}$Hf$_{19}$ sample over a rectangular area 90$\thinspace$nm by 45$\thinspace$nm in size, each scan taking 11 minutes to complete. Clear hopping behaviour is seen. The images in the movie all have a 3.0$\thinspace$nm vertical grey scale. All images were acquired at 20$\thinspace$pA tunnel current and 2.6$\thinspace$V sample bias. 

Movie two shows a sequence of scans on the Cu$_{50}$Hf$_{50}$ sample over an area of 45$\thinspace$nm by 22.5$\thinspace$nm, each taking 2 minutes 48 seconds to complete. Very fast noisy hopping is seen, and the overall movement of material is very much increased relative to the Cu$_{81}$Hf$_{19}$ film. The images in the movie all have a 2.5$\thinspace$nm grey scale. All images were acquired at 20$\thinspace$pA and 2.2 V.

Movie three shows the initial scanning sequence on the Cu$_{75}$Hf$_{25}$ sample. The first 9 frames are 45$\thinspace$nm in width and took 2 minutes 48 seconds for each frame. The remaining 16 frames are 90$\thinspace$nm in width and required 11 minutes to complete. The frame rate changes accordingly. All images were acquired at 19$\thinspace$pA and $2.2\thinspace$V with the exception of 5 of the larger frames, which used $2.1$, $2.0$, $1.9$, $1.7$ and $1.6\thinspace$V. These frames have a slightly noisier imaging state.

\subsection*{Reliable identification of hopping events in STM data}

Numerous sources of noise can complicate STM measurements. This is a major issue that must be confronted when analyzing STM noise patterns for physical meaning in dynamics. This is an emerging application of STM. The following extends the discussion of the identification of hopping in the main text. 

The primary requirement is to distinguish hopping events from the other sources of noise. Mechanical or electrical noise can contribute periodic oscillations that are easy to identify, but can also result in stochastic noise which might mimic fast hopping noise. More commonly, an unstable tip or incorrect feedback loop settings can also lead to stochastic noise that appears as short ``scars" in single scan lines, uncorrelated with features in neighbouring lines along the slow scan direction. Feedback settings may also contribute scars that are multiple scan lines wide when the tip scans over large topographic features; these are readily identified by comparison of the forward and backward scan images: feedback artifacts will appear on opposite sides of the topographic feature. Bearing these possibilities in mind, it is possible to unambiguously discriminate the portion of the hopping event distribution which has no overlap with other sources of noise in the measurement (see Figure 2 of the Supplementary Information). 
As detailed in the main text of the manuscript, a rigorous procedure is used to discriminate cluster hopping from scanning artifacts and noise. However, the implemented method does have the disadvantage of ignoring faster hops in analysis. Without further context, it is impossible to separate which fast scan noise represents hopping clusters. 

There is one situation within the present work for which some further context is available: imaging of the active surface after sputtering. Fast hopping is relevant in the frequency analysis of these data. In this case, imaging of sections previously scanned provides a control area within the scan, making it demonstrable that mechanical, electrical, or tip noise are not contributing to the pervasive high frequency noise seen elsewhere. The emergence of identifiable hopping as the high frequency noise disappears during repeated scanning also supports the conclusion that the high frequency noise is fast hopping. The frequency-based analysis applied to the argon-sputtered surface sequence cannot be applied to other data in the manuscript for two reasons: no control areas where noise density drops considerably are available, and dynamics are so much slower that the frequencies involved in hopping overlap the topographic frequencies in the power spectrum. 

\subsection*{Cross-checking for surface contamination}

Typically in STM experiments samples undergo extensive cleaning and annealing prior to measurements. This is impossible here because high temperature annealing would destroy the amorphous state and crystallize the film. Therefore, care must be taken to verify that the sample is clean and that observed structure and dynamics do not result from spurious contamination. In this experiment, surface dynamics were found to persist under all experimental conditions. The observations presented in the main article contribute significantly to ruling out contaminate contribution to the observed dynamics. Critically, the character and activity of the surface is consistent between observations made with two UHV STMs and three electron microscopes, independent of the transfer conditions used for the sample. The fact that the less surface sensitive SEM and TEM observations confirm the same behavior provides a clear, and robust verification that the cluster hopping and other dynamics are endemic to these amorphous samples. 

Additional tests were undertaken to verify that the samples were not contaminated or oxidized. The \em in situ \em removal of the surface by use of either the STM tip itself, or sputtering by argon ions did not eliminate the surface dynamics. 

Gentle heat treatments were used on all samples after after initial observations. Two thermal treatments were used. Three samples (Cu$_{81}$Hf$_{19}$, Cu$_{63}$Hf$_{37}$ and Cu$_{50}$Hf$_{50}$) were heated in UHV on the sample holder to $\sim$ 390 K for 2 hours. The other samples were subjected to a 48 h bake in the STM load lock at 400 K. Upon re-imaging following the heating, no significant change in structure or imaging was noted for any sample following heating. 

The electronic properties of a sample may be used as an indication of oxidation. A metallic surface and metallic tip have characteristic linear low bias current-voltage (I-V) relationships. All samples were tested for non-linearity in I-V curves at low bias voltages. Particular attention was applied to the Cu$_{85}$Hf$_{15}$ sample as it was the only sample exposed to air. All samples yielded linear responses.

Finally, some samples were coated with 10$\thinspace$nm of gold immediately ($\sim$ 5 seconds) after deposition of copper hafnium, without breaking high vacuum. Imaging these samples with the STM revealed no detectable surface dynamics (see Supplementary Information Figure 3). 

\subsection*{Identification of tip artifacts}

Numerous separate tips were utilized during the course of the experiments, in many cases specifically changing tips to rule out noise contributions from the tip in the observed dynamics. As discussed in the main text a detailed and reliable method of preparing tungsten tips in ultrahigh vacuum was employed. Typically we are able to prepare a stable tip that immediately yields atomic resolution on a crystalline surface. This also makes it very unlikely that the tips would consistently introduce a contaminant into the system. 

Beyond contributing stochastic noise, tip conditions can contribute many other effects which need to be identified and ruled out in STM data. Particularly important here are multiple tip effects, where more than one part of the tip contributes to the STM image. On rough surfaces, this is more likely. Most insidiously, it is possible to picture a single cluster being imaged by a second tip asperity and appearing to hop if that second tip is unstable. It is therefore critical to reject any data that has signs of multiple tips. These signs include double images, wispy features overlayed on the surface, or sudden disappearance of large, multi-cluster features. 

Tip interactions with the surface, such as tip crashes, adsorption onto or deposition of material from the tip, can lead to changes similar to the large hillock mergers described in this work. These events are identified through sudden changes in the imaging signal which correspond to physical changes in the tip. When material is adsorbed (or desorbed) from the tip, the feedback loop shifts to compensate for the longer (or shorter) tip. This shows up as a step in the slow scan direction when a permanent change to the tip is made, or as a transient scar when something adsorbs onto the tip and desorbs a short time later. Typically a tip change results in a change in the imaging resolution, noise level, or a vertical and/or lateral shift of the image as a different part of tip becomes dominant in imaging. Surface dynamics information was not extracted from images failing the preconditions for tip stability. This prescription was used to evaluate the dynamics seen on the surface and determine that tip contact does not play a major role in the dynamics and evolution of the CuHf surface. 

\subsection*{Temperature-dependence of hopping}

To probe the temperature dependence of the hopping dynamics, additional observation time was obtained on a second STM featuring variable temperature control (RHK UHV-3000). Using a fixed tunnelling current (150$\thinspace$pA), hopping activity observed on an Cu$_{85}$Hf$_{15}$ sample was found to strongly depend on temperature. Bias voltage was varied (1.0$\thinspace$V, 1.5$\thinspace$V, and 2.0$\thinspace$V) at each temperature. As with the measurements made on the Createc microscope, clusters appeared to jump, creating discontinuous topography from scan line to scan line. No bias voltage dependence was found. 

The number of scan lines for which a cluster is present, or absent from a particular location on the surface gives an indication of the hopping lifetime of clusters occupying that site. In order to extract this information from the image, a standard scar correction algorithm \cite{Gwyddion} can be used. Sudden jumps in data are detected as spikes which deviate from neighbouring points, both before and after, in the topographic derivative with respect to the slow scan direction. The distance between correlated jumps caused by the two transitions at the same site is extracted as the width of the scar identified by the algorithm. Analysis of the distribution of scar widths enabled an extraction of the lifetimes of hopping clusters for each image. High temperatures featured a much larger number of hopping sites, and even more prominently, much shorter hopping lifetimes. The fact that the copper hafnium surface is extremely active and the measured lifetimes reflect contributions from multiple cluster sizes prevents a simple application of an Arrhenius analysis to extract the energy barrier as has been noted in previous STM work applied to surface dynamics.\cite{PCL} The detection bandwidth available in STM scans is extremely limited. Consequently, the portion of the cluster population participating in observable hopping changes with temperature. This experiment provides valuable verification of the temperature dependent nature of the transitions.


\section*{Acknowledgements}

We are grateful for support from the Natural Science and Engineering Council of Canada, the Informatics Circle of Research Excellence, the National Institute for Nanotechnology, the Canadian Institute for Advanced Research, the Canada Research Chairs program, and Alberta Innovates. We would like to thank F.A. Hegmann for access to the Ultrafast Nanotools lab which was funded by the Canadian Foundation for Innovation. We also thank J. E. Losby and T. L. Cocker for useful discussions.

\section*{Author contributions statement}
J.A.J.B., C.M.B.H., D.M. and M.R.F.  conceived experiment(s), 
J.A.J.B., C.M.B.H., D.C.F., G.P., B.Z., and P.C. conducted experiment(s),
J.A.J.B. and E.J.L. analysed the results and
J.A.J.B., E.J.L., D.M. and M.R.F. prepared the manuscript.

\section*{Additional information}

\textbf{Competing financial interests:}  The authors declare no competing financial interests.

	\end{document}